\documentclass[intlimits,twoside,a4paper]{article}

  \usepackage{graphicx}        
\usepackage[cp1251]{inputenc}
\usepackage[eqsecnum]{cmpj3}

\articletype{Rapid Communication}

\issue{2026}{29}{2}{24601}
\doinumber{10.5488/CMP.29.24601}
\title[A link between the mesoscopic theory and the associative mean spherical approximation]%
{On the electrical double layer capacitance of the restricted primitive model: a link between the mesoscopic  theory and the associative mean spherical approximation}
\author[O. Patsahan]{O. Patsahan\orcid{0000-0002-5839-3893}
}
\address{
Yukhnovskii Institute for Condensed Matter Physics of the National
Academy of Sciences of Ukraine, 1~Svientsitskii Str., 79011 Lviv,
Ukraine
}
\date{Received 18 March 2026; revised 14 April 2026; accepted 17 April 2026; published 29 June 2026}

\begin{document}
	
	\maketitle
	
	\begin{abstract}
			The results for the electrical double layer capacitance and the density of ``free ions''  obtained from  the mesoscopic theory are compared with the corresponding results of the associative mean spherical approximation. While the first theory takes into account the  fluctuations of the charge density, the second theory assumes that the free ions and ion pairs  are in chemical  equilibrium according to the mass action law. Our results demonstrate a fairly good agreement between the two theories at high densities and low temperatures.
		\printkeywords  EDL capacitance,  concentrated electrolytes, association, mesoscopic theory, associative mean spherical approximation
		%
	\end{abstract}
	

The structure of the electrolyte/electrode interface plays a significant role in electrochemical processes such as electrochemical energy conversion, electrolysis, electrocatalysis, electrochemical devices, etc. This interface is particularly important for energy storage in electric double-layer (EDL) capacitors, or supercapacitors \cite{fedorov:14:0,Zhan2017,Kondrat2023}. 
 Understanding  the structure of EDLs at high ionic concentrations
is also important for fundamental reasons.

The study of ion distribution near charged surfaces has a long history and has been carried out using both theoretical and computer simulation methods.
The classical approach to electrolytes is the Debye-H\"{u}ckel theory, which is a linearized mean-field (MF) approximation \cite{barrat:03:0,fedorov:14:0}. The corresponding theory for the EDL is known as the Poisson--Boltzmann (PB) approximation. In this framework, ions are treated as  isolated point-like
charges in a solvent considered as a continuum dielectric. However, the classical description of EDLs is only valid for dilute
electrolytes but it is not suitable for concentrated electrolytes and room-temperature ionic liquids (RTIL).
Later, more sophisticated approaches, mostly based on the integral equation theories or different modifications of the PB equation, were developed~\cite{Blum1977,attard:93:0,Outhwaite1983,HOLOVKO1996,Henderson2009}.  
Initially, these approaches were applied to the simplest possible models, generally charged hard spheres in a uniform dielectric continuum (i. e., primitive models of electrolytes) next to a flat uniformly charged hard wall.

Recently, anomalous underscreening, i.e., unusually long-range screening lengths,
was experimentally observed for a number of concentrated electrolytes and RTIL \cite{smith:16:0,lee2017},  although, it was not detected in some  experimental studies \cite{Baimpos2014,kumar:22:0}. This phenomenon was extensively studied using theory and simulations~\cite{Kjellander2018,Rotenberg_2018,Adar2019,Cats2021,ciach:21:0,ciach:23:1,Safran2023,Patsahan2025,Coles2020,Hrtel2023,Yang2023,Elliott2024}. Unfortunately, different experimental techniques, approximate theories, and  simulations yield results that are not consistent with each other and the fundamental question of the structure and the screening length in dense ionic systems remains open.

In this letter, we do not have  space enough to
cover all theories that have been applied to the EDL. We only mention a few of them, in particular,
the mean spherical approximation (MSA)  applied to the EDL by Blum \cite{Blum1977},
the modified Poisson--Boltzmann  theory \cite{Outhwaite1983}, 
and the associative mean spherical approximation (AMSA) \cite{holovko1991effects,Blum95,Bernard96} applied for the description of an  electrified interface in \cite{HOLOVKO1996,Holovko2001}. 
The EDL capacitance 
was also studied using the classical density functional theory (DFT) (see e.g., \cite{Evans1980,Pizio2004,Jiang2014,Hrtel2017,Cats2022}). The DFT requires some approximate expression as
a starting point  for the intrinsic free-energy  functional of the  density profile and its implementation  is based on the variational principle. The latter makes it a computational tool to study
the EDL structure  of complex molecular systems.
Recently, the mesoscopic theory has been developed for the  EDL capacitance \cite{ciach:25:0,patsahan:26:0}. 
This theory goes beyond the MF approximation by taking into account the variance of the local charge that in concentrated inhomogeneous ionic systems is large.

The purpose of this letter is to compare the results for the EDL capacitance obtained within the mesoscopic theory  with the results of  the AMSA  which appears to be successful  for the description of ionic systems \cite{Jiang02,holovko2017application,Kalyuzhnyi2018}. 
To this end, we limited ourselves to the
restricted primitive model (RPM) at a flat electrode. In the RPM, spherical ions with
equal diameters $a_+=a_-=a$ and opposite charges are dissolved in a structureless
solvent characterized by the dielectric constant $\epsilon$. Below, we  present the main expressions for  the capacitance of the EDL obtained within the framework of both theories in the limit of the small voltage.

\paragraph{{\it EDL capacitance in the associative mean spherical approximation.}}
The AMSA is based on the theory of associating fluids.
In this theory, an ionic model is regarded to
be a mixture of free ions and ion pairs which are in chemical equilibrium
according to the mass action law (MAL).
In this approach, the differential capacitance at the small voltage has the form \cite{HOLOVKO1996}:
\begin{equation}
	C=\frac{\epsilon}{4\piup a}2\Gamma_{B}^*, \qquad \Gamma_{B}^*=\Gamma_{B}a,
	\label{C_AMSA}
\end{equation}
where screening parameter $\Gamma_{B}$ is calculated from the equation 
\begin{equation}
	4(\Gamma_{B}^*)^2(1+\Gamma_{B}^*)^3=\kappa^2(\alpha+\Gamma_{B}^*).
	\label{Gamma_B}
\end{equation}
In (\ref{Gamma_B}), $\kappa$ is the inverse Debye screening length in $1/a$ units and $\alpha$ is the degree of dissociation which satisfies the MAL:
\begin{equation}
	1-\alpha=\frac{\rho}{2}\alpha^2 K,
	\label{alpha}
\end{equation}
where $\rho=\rho_++\rho_-$ is the dimensionless density of ions. In (\ref{alpha}),   $K=K_{\text{ass}}^{(0)}K^{\gamma}$ is the association constant where  the thermodynamic association constant $K_{\text{ass}}^{(0)}$ is the infinite-dilute limit of $K$. $K^{\gamma}$ is the concentration-dependent part which is determined as the ratio of the activity coefficients of free ions to the activity coefficient of the ion pair. In the AMSA, there is a certain kind of arbitrariness in the definition of the ion pair and hence the thermodynamic association constant $K_{\text{ass}}^{(0)}$. We choose $K_{\text{ass}}^{(0)}$ in the form introduced by Ebeling \cite{Ebeling1968}.
Ebeling's definition of the ion association constant together with the MSA contribution gives an exact second ionic virial coefficient. 

$K^{\gamma}$ is given by
\[
K^{\gamma}=g^{\text{hs}}(r=a)\exp\Bigg[-l_{\text B}\frac{\Gamma_{B}^*(2+\Gamma_{B}^*)}{(1+\Gamma_{B}^*)^2}\Bigg],
\]
where  $l_{\text B}=\beta e^2/(a\epsilon) $ is the Bjerrum length in $a$ units and
\[
g^{\text{hs}}(r=a)=\frac{1-\eta/2}{(1-\eta)^3}, \qquad \eta=\frac{\piup}{6}\rho, 
\]
is the contact value of the pair distribution function between
the hard-sphere fluid particles of diameter $a$. It is worth noting that $\rho_{\rm{free}}=\alpha\rho$ determines the free ion density.

Without association ($\alpha=1$), $\Gamma_B$ reduces to the screening parameter $\Gamma^*=\Gamma a$ in the MSA
\begin{equation}
	\Gamma^*=\frac{1}{2}\Big(\sqrt{1+2\kappa}-1\Big).
	\label{Gamma_MSA}	
\end{equation}
In the MSA,  the differential capacitance  has the form (\ref{C_AMSA}) when   $\Gamma_B^*$  is replaced with  $\Gamma^*$.

\paragraph{{\it EDL capacitance in the mesoscopic theory.}}
Within the framework of the mesoscopic theory, very simple expressions for the EDL capacitance in dilute and concentrated electrolytes were obtained \cite{ciach:25:0,patsahan:26:0}. These expressions  for the RPM near a flat metallic electrode are considerably different. They become identical, however, at the Kirkwood line \cite{kirkwood:36:0}  separating the monotonous and
oscillatory asymptotic decays of the charge density. 
The derivation of the formalism of the theory is described in  detail in \cite{ciach:18:1,ciach:25:0,patsahan:26:0}.

The  charge-density profile on the small-density side of the Kirkwood line  has the form:
 \[
 c(z)=A_1\Big(\mathrm{e}^{-a_1 z}-\frac{a_2}{a_1} \mathrm{e}^{-a_2z}\Big),
 \]
 where $a_1$, $a_2$ are in $1/a$ units and $a_1>a_2$. For the small voltage,  one gets for the capacitance 
\begin{equation}
C=\frac{\epsilon a_2(1-\delta)}{4\piup a (1-\frac{a_2}{a_1})}\simeq \frac{\epsilon a_2}{4\piup a}\simeq \frac{\epsilon }{4\piup \lambda_{\mathrm{D}}},
\label{C_Debye}
\end{equation}
where, for dilute electrolytes, $\delta= 1-\mathrm{e}^{-a_2/2}+\mathrm{e}^{-a_1/2}\simeq a_2/2+\mathrm{e}^{-a_1/2}\ll 1$,  $1/a_2\rightarrow\lambda_{\mathrm{D}}$~\cite{leote:94:0,ciach:03:1} and $\lambda_{\mathrm{D}}$ is the Debye screening length in $a$ units.

On the large-density side of the Kirkwood line, $c(z)$ decays in an oscillatory way, 
\[
c(z)=A_2\exp(-\alpha_0 z)\sin(\alpha_1 z +\theta),
\]
where  $\alpha_0$ and  $\alpha_1$ are  obtained from the pole analysis of the Fourier transform of the charge-charge correlation functions $\langle\tilde{c}(\mathbf{k})\tilde{c}(-\mathbf{k})\rangle$ extended
to the complex $k$ plane.

The EDL capacitance  for the RPM   takes, in the limit of vanishing
voltage, the simple form:
\begin{equation}
	\label{Cd}
	C=\frac{\epsilon(\alpha_0^2+\alpha_1^2)}{4\piup a }f,
\end{equation}
where the factor resulting from the mesoscopic charge distribution at the electrode reads \cite{patsahan:26:0}
\begin{equation}
	\label{f}
	f=\frac{\sin(\alpha_1/2)\exp(-\alpha_0/2)}{\alpha_1}.
\end{equation}

In the mesoscopic theory, $\langle\tilde{c}(\mathbf{k})\tilde{c}(-\mathbf{k})\rangle$ takes the form:
\begin{equation}
	\label{Ccc4}
\langle\tilde{c}(\mathbf{k})\tilde{c}(-\mathbf{k})\rangle=\Bigg( \beta\tilde V_{\text{C}}(k)+\frac{1}{\rho_{\mathrm{R}}}\Bigg)^{-1},
\end{equation}
where $ \beta=1/(k_{\text B}T)$, and $ \beta V_{\text{C}}(r)=l_{\text B} \theta(r-1)/r$ is the Coulomb potential truncated for distances smaller than the ionic diameter  (the unit step function is $\theta(r-1)=1$ for $r>1$ and $\theta(r-1)=0$ for $r<1$).  $l_{\text B}$ is the Bjerrum length in $a$ units.
The Fourier transform $ \beta\tilde V_{\text{C}}(k)$ takes a negative minimum for $k=k_0>0$. The ``effective'' density
$\rho_R$ satisfies,  in the Gaussian approximation,  the equation~\cite{ciach:23:1}
\begin{equation}
	\label{roR}
	\frac{1}{\rho_{\mathrm{R}}}=\frac{1}{\rho}+\frac{\langle c^2\rangle}{\rho^3},
\end{equation}
where $\rho=\rho_++\rho_-$ and $\langle c^2\rangle$ is the variance of the local charge  given by
\begin{equation}
	\label{<phi2>}
	\langle c^2\rangle\approx \int \frac{\rd{\bf k}}{(2\piup)^{3}}\langle\tilde{c}(\mathbf{k})\tilde{c}(-\mathbf{k})\rangle.
\end{equation}
Equations~(\ref{Ccc4})--(\ref{<phi2>}) should be solved self-consistently.

 In dense electrolytes $\rho_R<\rho$ due to spontaneous formation of oppositely charged neighboring regions with the size $\sim a$. In turn,  it means that  aggregates  are formed with oppositely charged nearest neighbors and fewer free ions remain in the system. Therefore,  $\rho_R$ is related to the number density of  free ions.

\begin{figure}[h]
	\begin{center}
		\includegraphics[width=0.40\textwidth,angle=0,clip=true]{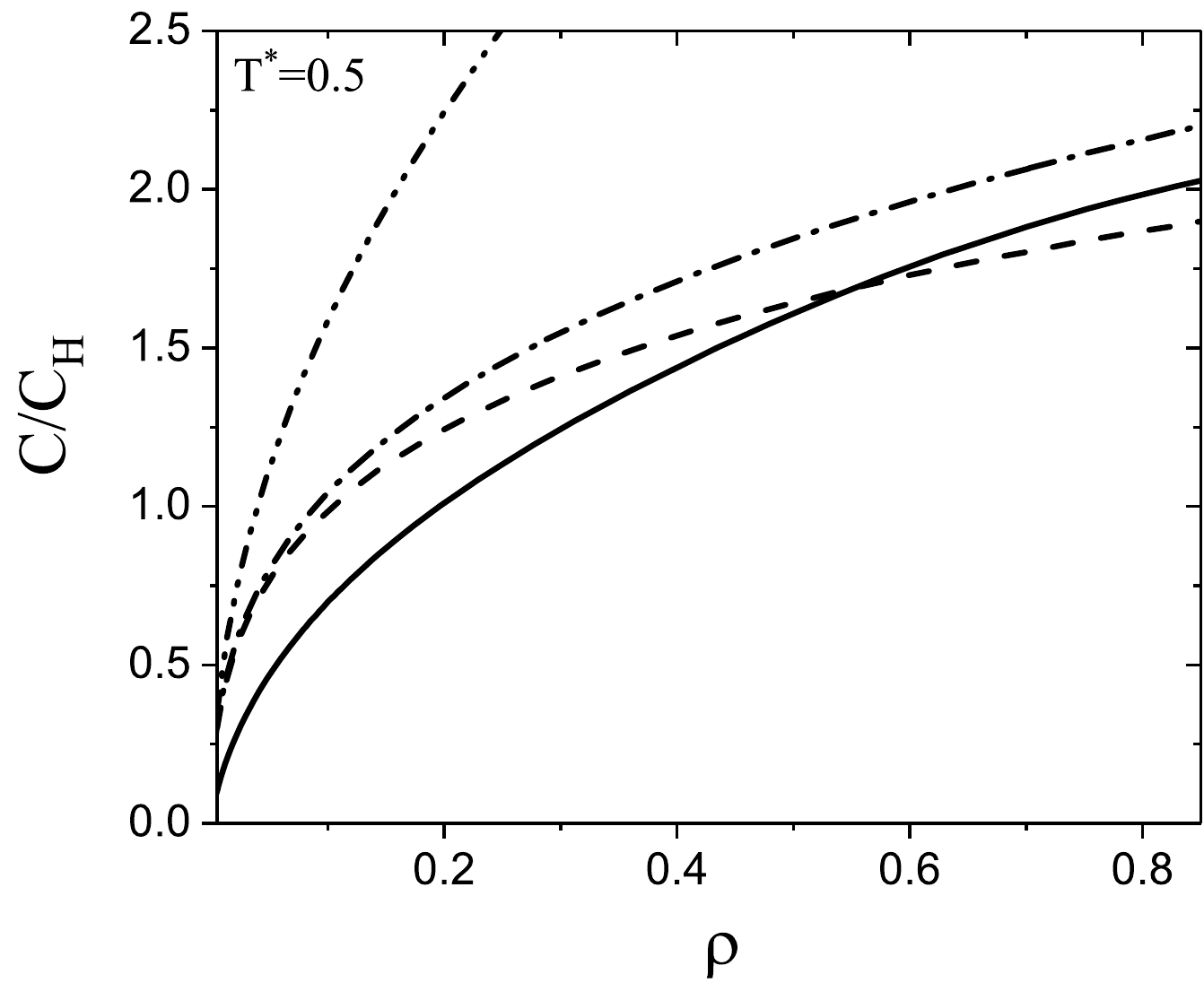}
		\includegraphics[width=0.41\textwidth,angle=0,clip=true]{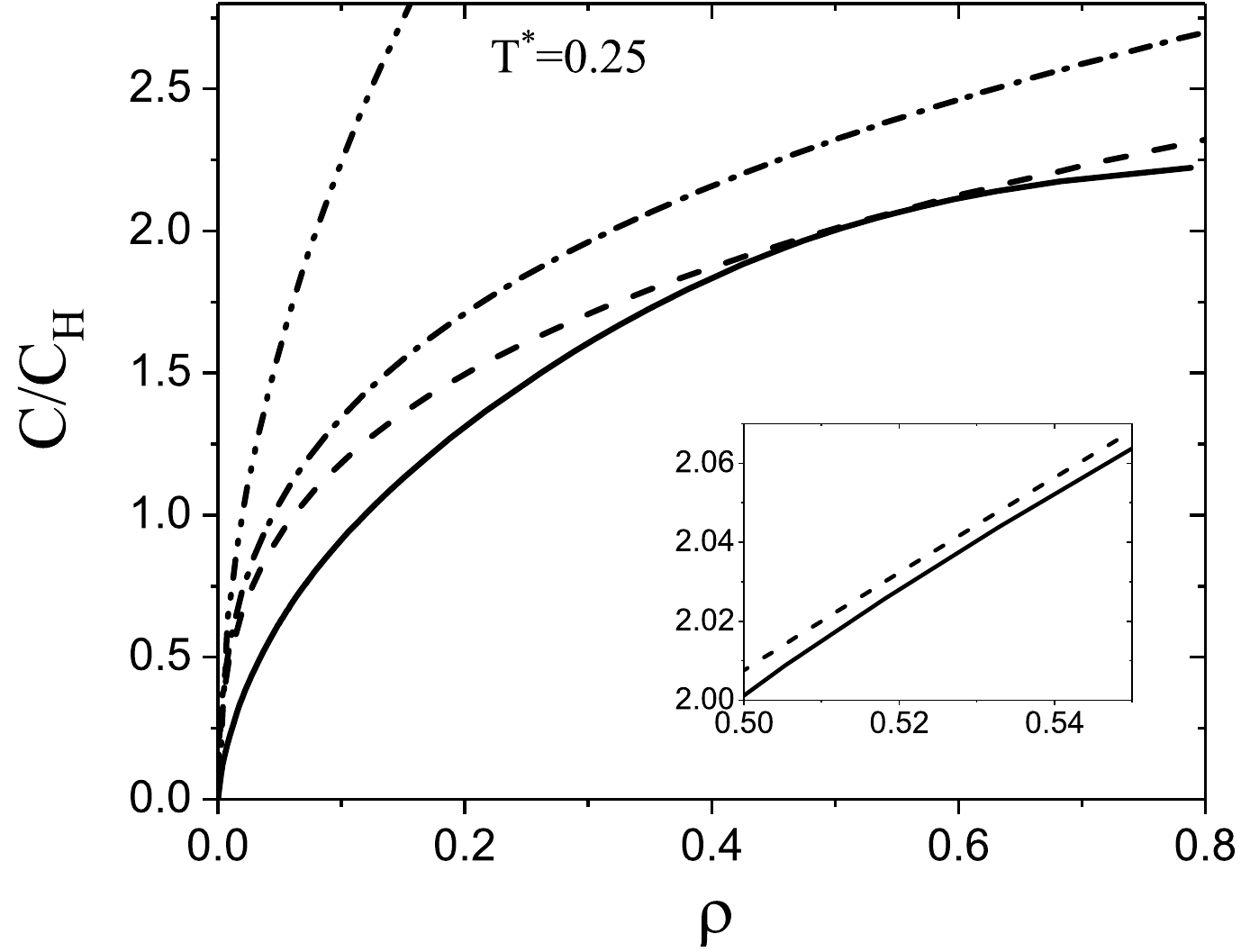}
		\caption{The EDL capacitance $C$ of the RPM in units of the Helmholtz capacitance $C_{\mathrm{H}}=\epsilon/(4\piup a)$ obtained from the mesoscopic theory (solid line), MSA (dash-dotted line), AMSA with Ebeling's association constant (dashed line) for  $T^*=0.5$ (left-hand  panel) and $T^*=0.25$ (right-hand panel). The Debye capacitance, equation~(\ref{C_Debye}), is shown by a dash-dot-dotted line.  $\rho$ is the dimensionless density of ions, $T^*=1/l_{\text B}$,  $l_{\text B}$ is the Bjerrum length in $a$ units ($a$ is the ion diameter). The inset in the right-hand panel shows the magnified plot of  the range $0.5<\rho<0.55$ where the capacitance obtained from the mesoscopic theory and from the AMSA takes almost the same values.
		}\label{C}
	\end{center}
\end{figure}

\begin{figure}[h]
	\begin{center}
		\includegraphics[width=0.40\textwidth,angle=0,clip=true]{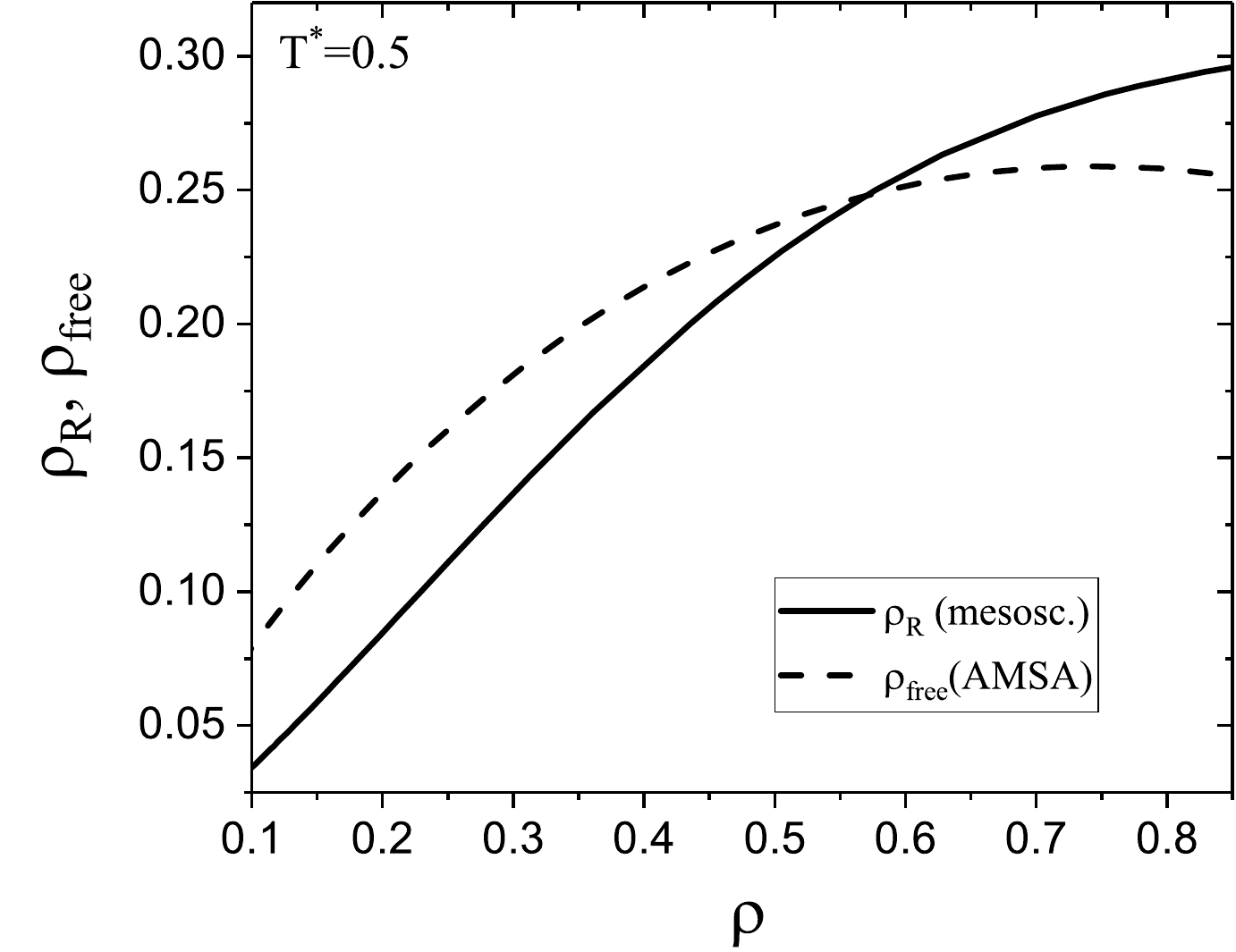}
		\includegraphics[width=0.41\textwidth,angle=0,clip=true]{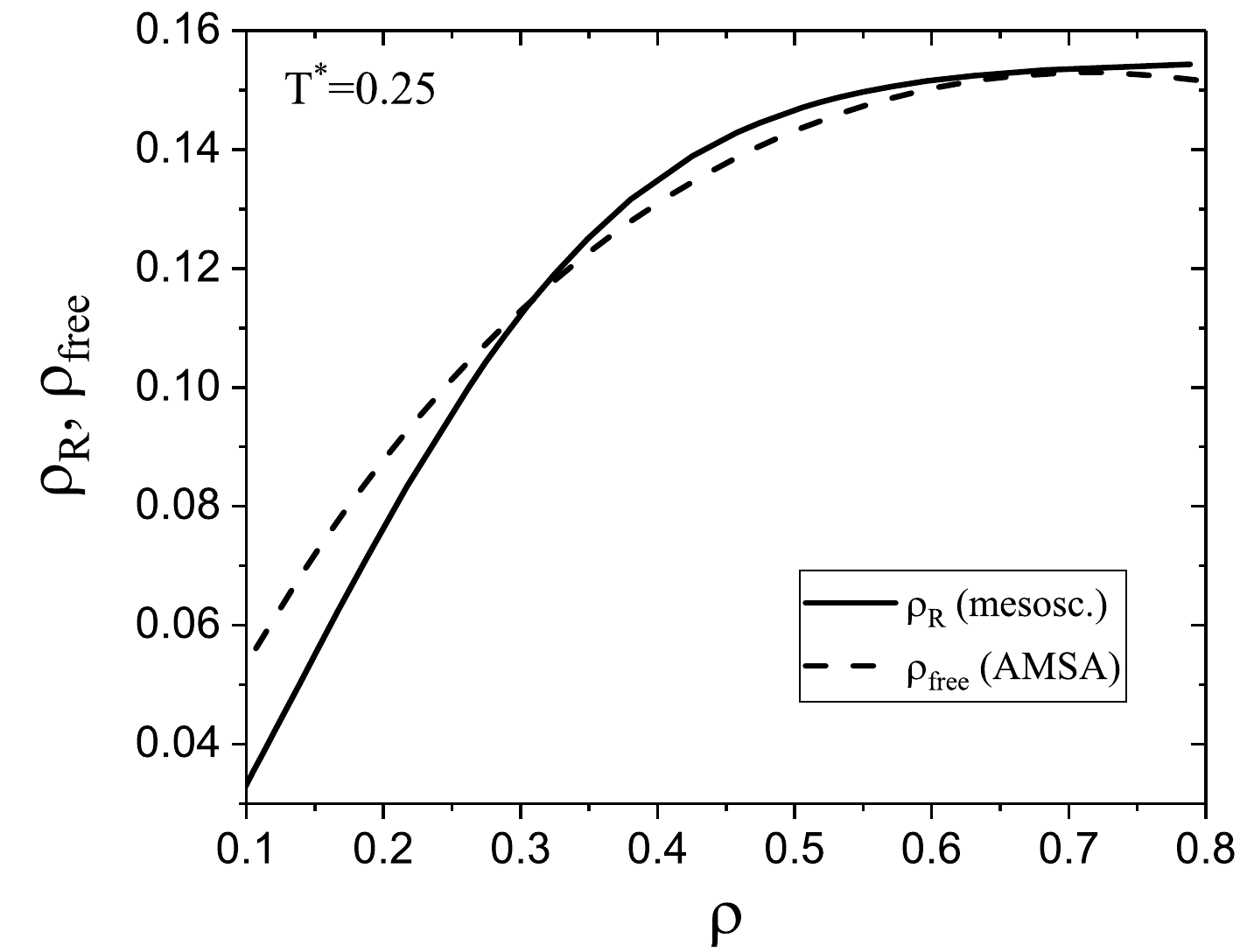}
		\caption{Effective density of ions $\rho_R$ obtained from the mesoscopic theory, equation (\ref{roR}), (solid line) and the density of free ions $\rho_{{\rm free}}=\alpha\rho$ ($\alpha$ is the degree of dissociation) obtained from the AMSA with Ebeling's association constant (dashed line) for  $T^*=0.5$ (left-hand panel) and $T^*=0.25$ (right-hand panel).  $\rho$ is the dimensionless density of ions, $T^*=1/l_{\text B}$, $l_{\text B}$ is the Bjerrum length in $a$ units ($a$ is the ion diameter).  
		}
		\label{rho_R}
	\end{center}
\end{figure}


\paragraph{{\it Results.}}
Using the above-mentioned equations we calculated the EDL capacitance $C$ at small voltage  as a function of the bulk number density of ions $\rho$ for the fixed dimensionless temperature $T^*=1/l_{\text B}$.
In figure~\ref{C}, the  capacitance $C$  in units of the Helmholtz capacitance $C_{\mathrm{H}}=\epsilon/(4\piup a)$ obtained from the mesoscopic theory, MSA and AMSA is shown as a function of the bulk number density of ions  for  $T^*=0.5$ and $T^*=0.25$. The $\rho$-dependence of the Debye capacitance, equation~(\ref{C_Debye}), is also shown for comparison. It is seen that the results obtained from the mesoscopic theory are rather in good agreement  with the AMSA results when  $\rho$ increases. Moreover, for $T^*=0.25$ and $\rho\geqslant 0.4$ the results obtained from both theories  are very close and almost coincide for $0.5<\rho<0.55$ (see the inset in the right-hand panel).

Besides the  capacitance curves, it is also interesting to compare the ``effective'' density of ions $\rho_R$ in the mesoscopic theory given by equations (\ref{roR})--(\ref{<phi2>})  with the density of free ions $\rho_{\rm{free}}=\alpha\rho$ ($\alpha$  and $\rho$ are  the dissociation constant and the number density of ions, respectively) obtained within the framework of the AMSA theory [see equations (\ref{Gamma_B})--(\ref{alpha})]. The corresponding dependence of $\rho_R$ and $\rho_{{\rm free}}$ on $\rho$ is shown in figure~\ref{rho_R} for two values of the reduced temperature. It is clearly seen that the both curves are very close to each other for $T^*=0.25$ and $\rho\geqslant 0.3$ (figure~\ref{rho_R}, right-hand panel).

\paragraph{{\it Conclusions.}}
We  compared the results for the EDL capacitance and the density of free ions obtained from two different theories, namely the associative mean spherical approximation (AMSA) and the mesoscopic theory. The first theory  assumes that the free ions and ion pairs in the ionic system are in chemical  equilibrium according to the MAL, whereas the second theory takes into account the  fluctuations in the local charge density.

First, we compared the results for the EDL capacitance in the limit of small voltage as a function of the bulk number density of ions $\rho$ for two values of the reduced temperature, $T^* = 0.5$ and $0.25$, where $T^*$ is the inverse reduced Bjerrum length. Overall, the results demonstrate fairly good agreement at higher densities, and the agreement is better at lower temperatures.

We also compared the ``effective'' density of ions $\rho_R<\rho$, which appeared in the mesoscopic theory as a result of fluctuations being taken into account, with the density of free ions $\rho_{\rm{free}}=\alpha\rho$ in the AMSA where $\alpha$ is the degree of dissociation. We obtained a fairly good agreement between the theories  for $T^*=0.25$ in the range of the reduced density  $\rho\geqslant 0.3$.

Our result suggests that the self-consistent Gaussian approximation is equivalent to the equilibrium between free ions and ion pairs. For large density of ions, however, larger neutral clusters can also be formed. 
We can suppose that taking into account the equilibrium between different aggregates would correspond to the mesoscopic theory beyond the  Gaussian approximation.  
One should  also note that the AMSA  modified  by including ion trimers and tetramers satisfactorily reproduces the thermodynamic measurement data up to high concentrations for nonaqueous solutions with solvents of lower permittivity~\cite{Barthel2000,Holovko2002}.
 It will be interesting to compare the results obtained from both theories accounting for the equilibrium between larger clusters in a future work.

\bibliographystyle{cmpj} 
\bibliography{bibliography_CMP_2026}


\ukrainianpart

\title{Щодо ємності подвійного електричного шару обмеженої примітивної моделі:  зв'язок між мезоскопічною теорією та асоціативним середньосферичним наближенням}
\author{О. Пацаган}
\address{Інститут фiзики конденсованих систем імені І. Р. Юхновського НАН України, вул.~Свенцiцького, 1, 79011 Львiв, Україна
}

\makeukrtitle

\begin{abstract}
	\tolerance=3000%
Результати для ємності подвійного електричного шару та густини ``вільних іонів'', отримані з використанням мезоскопічної теорії, порівнюються з відповідними результатами асоціативного середньосферичного наближення. У той час як перша теорія враховує флуктуації густини заряду, друга теорія припускає, що вільні іони та іонні пари знаходяться в хімічній рівновазі згідно із законом діючих мас. Наші результати демонструють досить добру узгодженість між двома теоріями при високих густинах та низьких температурах.
	\keywords ємність подвійного електричного шару, концентровані електроліти, асоціація,  мезоскопічна теорія, асоціативне середньосферичне наближення
\end{abstract}

\end{document}